\documentstyle[12pt]{article}
\begin{document}
\title{STRUCTURE OF AN ELASTIC LATTICE PINNED BY HOPPING DEFECTS}
\author{Eugene M. Chudnovsky\\ Physics Department, CUNY Lehman College\\
        Bedford Park Boulevard West, Bronx, New York 10468-1589\\}
\date{11 March 1998}
\maketitle
\begin{abstract}
The translational order in a three-dimensional lattice pinned by hopping 
defects is studied. It is suggested that the equilibrium state of the
lattice crosses from a glass to a crystal on the rate of hopping.  It 
is argued that such a transition may exist in flux lattices of 
high-temperature superconductors due to the hopping of oxygen.\\  
\end{abstract}
\vspace{2em}
\vspace{2em}

It is known that below four dimensions the translational order 
in a lattice subject to a static random background (pinning) is 
destroyed by a however weak interaction with the background 
\cite{Larkin,Imry,Brandt1,Review}. This fact has received much
attention mostly in connection with pinning of flux lattices in
superconductors, where translational correlations determine such
important quantities as the resistivity and the critical current.
The problem has been re-examined lately on the issue of the explicit 
form of the correlation function at large distances 
\cite{Nattermann,Bouchaud,Giamarchi,Carraro,Carpentier,CD}. 
Apart from that question, 
it is generally agreed that the correlation length of the 
translational order scales with the strength of pinning $f$ as
$1/f^{2/(4-d)}$ \cite{Larkin,Imry}. Thus, pinning, no matter how
weak, leads to a glass state \cite{Fisher} characterized by short range 
translational correlations. This effect of the static disorder 
is very different from the effect of the thermal disorder, as one 
can see from the fact that a finite temperature is needed  to melt 
the lattice in three dimensions. The melting of the flux-line 
lattice has been intensively discussed in connection with the 
irreversibility line in the phase diagram of high temperature 
superconductors \cite{Nelson,Houghton,Brandt2,Feigel'man,Review}.   
The question addressed by this paper is what happens if the random 
background potential, static in the first approximation, is allowed
to fluctuate slowly in time due to the thermal hopping of pinning 
centers. We shall study this problem for a three-dimensional triangular 
lattice with a dissipative dynamics, weakly pinned by hopping point defects.

The difference between static and thermal disorder in three dimensions 
stems from the fact that the statistical average of the static
random pinning force over a large distance grows faster with the distance
than the statistical effect of the random force due to thermal fluctuations.
The hopping of pinning centers softens, through time average,  this 
devastating large-distance effect of the static disorder. At high hopping 
rate its effect on the elastic lattice is equivalent to the thermal noise. 
We will argue that the translational order 
in the lattice should be restored when the rate of hopping attains some 
critical value and will discuss possible implications of that for 
superconductors. This phenomenon could also occur
in  charge density waves  \cite{Dai} and Wigner crystals 
\cite{Andrei} in semiconductors, atomic monolayers on imperfect
surfaces \cite{Nagler}, and other lattice systems where hopping of 
pinning centers may exist.  

In a continuous approach, deformations of the lattice are described by 
the displacement field ${\bf u}(\bf r)$. The free energy of the system 
is \cite{Landau}
\begin{equation}
F=\frac{1}{2}{\int}\,d^{3}r\,[(C_{11}-C_{66})({\partial}_{\alpha}
u_{\alpha})^{2}+C_{66}({\partial}_{\alpha}u_{\beta})^{2}+C_{44}
({\partial}_{z}u_{\alpha})^{2}]-{\int}\,d^{3}ru_{\alpha}f_{\alpha}\;\;\;,
\end{equation}
where ${\alpha},{\beta}$ can be $x$ or $y$; $C_{11}, C_{44}$, and $C_{66}$ 
are the elastic moduli of the triangular lattice; ${\bf f}({\bf r},t)$ is the
force associated with random pinning. It should be noted that the
replacement of the random potential by
the random force term in Eq.(1) is only true for small displacements as it  
fails to account properly for the periodicity of the lattice and the potential
nature of the interaction with the random background
\cite{Nattermann,Bouchaud,Giamarchi,Carraro,Carpentier,CD}. 
For our purpose, however,
the random force model should be sufficient because it gives the correct
estimate of the correlation length. The static Gaussian noise is equivalent
to $<f_{\alpha}({\bf q})f_{\beta}({\bf q}')>=(2{\pi})^{3}W{\delta}_
{{\alpha}{\beta}}{\delta}({\bf q}+{\bf q}')$ for the Fourier transform of
${\bf f}({\bf r})$. Here $W$ characterizes the strength of pinning. The
hopping of pinning centers can be introduced by 
assuming that $<f_{\alpha}({\bf q},t)f_{\beta}({\bf q}',t')>$ is proportional 
to ${\exp}(-{\Gamma}|t-t'|)$, where ${\Gamma}$ is the rate of hopping. Since
pinning sites are independent, both $W$ and ${\Gamma}$ should be independent of 
${\bf q}$ \cite{dispersion}. This gives
\begin{equation}
<f_{\alpha}({\bf q},{\omega})f_{\beta}({\bf q}',{\omega}')>=
(2{\pi})^{4}\frac{2{\Gamma}W}{{\Gamma}^{2}+{\omega}^{2}}{\delta}_
{{\alpha}{\beta}}{\delta}({\bf q}+{\bf q}'){\delta}({\omega}+{\omega}')\;\;\;.
\end{equation}

At not very low temperatures the dynamics of the flux-line lattice is 
dissipative \cite{Review},
\begin{equation}
{\eta}\frac{{\partial}{\bf u}}{{\partial}t}=
-\frac{{\delta}F}{{\delta}{\bf u}}\;\;\;,
\end{equation}
with $F$ given by Eq.(1) and ${\eta}$ being the drag coefficient. 
Substituting into this equation
\begin{equation}
{\bf u}({\bf r}, t)={\int}\frac{d^{3}q\,d{\omega}}{(2{\pi})^{4}}
{\bf u}({\bf q}, {\omega})e^{i{\omega}t-i{\bf q}{\cdot}{\bf r}}\;\;\;,
\end{equation}
one obtains for the Fourier transform of ${\bf u}$ 
\begin{equation}
i{\omega}{\eta}u_{\alpha}+(C_{11}-C_{66})({\bf q}{\cdot}{\bf u})q_{\alpha}
+(C_{66}q_{\perp}^{2}+C_{44}q_{z}^{2})u_{\alpha}=
f_{\alpha}({\bf q}, {\omega})\;\;\;,
\end{equation}
where the elastic moduli, in general, 
depend on ${\bf q}$;$\;\;\;q_{\perp}^{2}=q_{x}^{2}+q_{y}^{2}$. The solution 
of this equation is
\begin{equation}
u_{\alpha}({\bf q}, {\omega})=(i{\omega}{\eta}+C_{11}q_{\perp}^{2}+
C_{44}q_{z}^{2})^{-1}\frac{({\bf q}{\cdot}{\bf f})q_{\alpha}}{q^{2}_{\perp}}
+(i{\omega}{\eta}+C_{66}q_{\perp}^{2}+C_{44}q_{z}^{2})^{-1}\left[f_{\alpha}-
\frac{({\bf q}{\cdot}{\bf f})q_{\alpha}}{q_{\perp}^{2}}\right]\;\;\;.
\end{equation}
To determine the correlation length we shall compute 
\begin{equation}
B({\bf r})=\frac{1}{2a^{2}}<[{\bf u}({\bf r},t)-{\bf u}(0,t)]^{2}>\;\;\;,
\end{equation}
where $a$ is the lattice parameter. This can be accomplished by substituting 
here Eq.(4) with ${\bf u}({\bf q}, {\omega})$ given by Eq.(6), and averaging 
over random forces with the help of Eq.(2). Working out delta-functions one
obtains
\begin{equation}
B({\bf r})=\frac{W}{2a^{2}}{\int}\frac{d^{3}q\,d{\omega}}{(2{\pi})^{4}}\frac{2{\Gamma}}
{{\Gamma}^{2}+{\omega}^{2}}[1-{\cos}({\bf q}{\cdot}{\bf r})]\left[\frac{1}
{{\omega}^{2}{\eta}^{2}+(C_{11}q_{\perp}^{2}+C_{44}q_{z}^{2})^{2}}+\frac
{1}{{\omega}^{2}{\eta}^{2}+(C_{66}q_{\perp}^{2}+C_{44}q_{z}^{2})^{2}}
\right]\;\;.
\end{equation}
Further integration over ${\omega}$ gives
\begin{eqnarray}
B({\bf r}) & = & \frac{W}{2a^{2}}{\int}\frac{d^{3}q}{(2{\pi})^{3}}
[1-{\cos}({\bf q}{\cdot}{\bf r})]
[({\Gamma}{\eta}+C_{11}q_{\perp}^{2}+C_{44}q_{z}^{2})^{-1}
(C_{11}q_{\perp}^{2}+C_{44}q_{z}^{2})^{-1} \nonumber \\
           & + & ({\Gamma}{\eta}+C_{66}q_{\perp}^{2}
+C_{44}q_{z}^{2})^{-1}(C_{66}q_{\perp}^{2}+C_{44}q_{z}^{2})^{-1}]\;\;\;.
\end{eqnarray}
The effect of hopping now becomes apparent. It softens the small $q$ 
singularity under the integral. When ${\Gamma}$ goes to zero, 
Eq.(9) reduces to the well-known expression for $B$ in the case of 
a static disorder \cite{Larkin,Brandt1}.  

We shall now perform the final integration over ${\bf q}$ in Eq.(9).
At small ${\Gamma}$ the integral is dominated by small $q$. To shorten 
the formulas we shall assume that $C_{11}>>C_{66}$, which is true for flux 
lattices at small $q$ \cite{Remark}. Introducing  
\begin{eqnarray}
R & = & \left(x^{2}+y^{2}+\frac{C_{66}}{C_{44}}z^{2}\right)^{1/2} 
\nonumber \\
R_{0} & = & 16{\pi}a^{2}C_{44}^{1/2}C_{66}^{3/2}/W \nonumber \\
R_{h} & = & \left(\frac{C_{66}}{{\Gamma}{\eta}}\right)^{1/2}
\end{eqnarray}
and rescaling variables under the integral, one obtains after a series
of standard integrations
\begin{equation}
B(R)=2\frac{R_{h}}{R_{0}}\left\{1-\frac{R_{h}}{R}\left[1-
{\exp}\left(-\frac{R}{R_{h}}\right)\right]\right\}\;\;\;.
\end{equation}
Here $R_{0}$ is the Larkin-Ovchinnikov correlation length \cite{Larkin}
due to the static disorder, while $R_{h}$ is the length of the diffusion
of the unpinned flux-line lattice during the time ${\Gamma}^{-1}$.

Formally, the translational correlation function given by Eq.(11) diverges
at $R{\rightarrow}\infty$ only when $R_{h}{\rightarrow}\infty$, that is,
at ${\Gamma}{\rightarrow}0$. In that limit Eq.(11) coincides with the
Larkin's result, $B=R/R_{0}$. At any non-zero ${\Gamma}$, $\;\;\;B$ approaches
a constant $2R_{h}/R_{0}$ at $R{\rightarrow}\infty$. This result should be
taken with caution, however. Indeed, Eq.(11) does not provide the correct 
answer for translational correlations at large distances as
it has been derived in the random force approximation, without taking into
account the periodicity of the lattice. The answer for the random potential
problem is not known. It is believed, however, that the random force
approximation provides a good estimate of the translational correlation length,
$R_{t}$. The latter is the distance at which the sites of the deformed 
lattice can no longer sustain one-to-one correspondence with the sites of
the perfect lattice.   It can be obtained as a solution of the equation 
$B(R_{t})=1$. At ${\Gamma}{\rightarrow}0$, $\;\;\;B=R/R_{0}$, and, thus, 
$R_{t}=R_{0}$. As ${\Gamma}$ increases, $R_{t}=R_{0}$ remains a good 
approximation at $R_{h}>>R_{0}$. The significant change in $R_{t}$ occurs 
when $R_{h}$ becomes comparable with $R_{0}$. The translational correlation 
length grows rapidly as $R_{h}$ decreases below $R_{0}$ and becomes infinite 
at $R_{h}=R_{0}/2$. One should expect, therefore, that the long range 
translational order becomes restored at
\begin{equation}
{\Gamma}_{c}=\frac{4C_{66}}{{\eta}R_{0}^{2}}
\end{equation}

The limit of $R_{h}$ large compared to $R_{0}$ corresponds to small 
${\Gamma}$ and/or small ${\eta}$ \cite{damping}. In that limit the lattice 
adjusts to the evolution of the pinning potential fast enough to 
make this situation similar to 
the case of the static disorder. In the opposite limit of $R_{h}$ small 
compared to $R_{0}$, the pinning potential changes fast enough to
make its effect on the lattice similar to the effect of thermal fluctuations.
The latter effect is equivalent, through the Einstein relation, to the random 
force with the correlation function
\begin{equation}
<f_{\alpha}({\bf q},{\omega})f_{\beta}({\bf q}',{\omega}')>=
(2{\pi})^{4}2T{\eta}{\delta}_
{{\alpha}{\beta}}{\delta}({\bf q}+{\bf q}'){\delta}({\omega}+{\omega}')\;\;\;.
\end{equation} 
Repeating all steps leading to Eq.(9), one obtains  
\begin{equation}
B({\bf r})=\frac{T}{2a^{2}}{\int}\frac{d^{3}q}{(2{\pi})^{3}}
[1-{\cos}({\bf q}{\cdot}{\bf r})]\left[\frac{1}
{C_{11}q_{\perp}^{2}+C_{44}q_{z}^{2}}+\frac
{1}{C_{66}q_{\perp}^{2}+C_{44}q_{z}^{2}}
\right]
\end{equation} 
for the thermal disorder. This expression is finite at all $R$ and allows
to estimate the melting temperature through the Lindemann criterion 
\cite{Nelson,Houghton,Brandt2,Feigel'man,Review}, $B(a)=c^{2}_{L}$,
where $c_{L}\,<\,1$ is the Lindemann number. 
Equations (9) and (14) become identical in the limit of large ${\Gamma}\;\;$ 
(${\Gamma}\;>>\,C_{66}q_{BZ}^{2}/{\eta}$) if one replaces $W/{\Gamma}$ 
in Eq.(9) by ${\eta}T$, which represents the transition from hopping to 
thermal noise. Here $q_{BZ}=(8{\pi})^{1/2}/3^{1/4}a$ is the radius of the
Brillouin zone. It is easy to see that as long as $R_{0}$ remains large
compared to $a$, that is, in the weak pinning regime, ${\Gamma}_{c}$ of 
Eq.(12) is always small compared to the hopping rate, 
${\Gamma}\;{\sim}\;C_{66}q_{BZ}^{2}/{\eta}$, at which hopping crosses to
the thermal noise.  

For a superconductor $C_{66}=B_{c2}B{\xi}^{2}/32{\pi}{\lambda}^{2}$ 
\cite{Larkin,Brandt1}, where ${\xi}$ and ${\lambda}$ are the coherence length
and the penetration length respectively. The dissipation coefficient 
${\eta}$ equals \cite{Bardeen} $B_{c2}{\Phi}_{0}/{\rho}_{n}c^{2}a^{2}$,
where ${\Phi}_{0}$ is the flux quantum, 
${\rho}_{n}=4{\pi}{\lambda}^{2}{\nu}_{e}/c^{2}$ is the normal state
resistivity (${\nu}_{e}$ being the normal electron collision frequency), and
$a=(2^{1/2}/3^{1/4})({\Phi}_{0}/B)^{1/2}$ is the flux lattice spacing.
Substituting all that into Eq.(12), one obtains a remarkably simple expression
for ${\Gamma}_{c}$:
\begin{equation}
{\Gamma}_{c}=\frac{{\nu}_{e}}{\sqrt{3}}\left(\frac{\xi}{R_{0}}\right)^{2}\;\;\;.
\end{equation}

Let us now study the question whether the corresponding rate of hopping
may exist in superconductors. Measurements of translational correlations
in YBCO and BSCCO \cite{Gammel,Murray} suggest that at a low field the
ratio ${\xi}/R_{0}$ may be as small as $10^{-3}$. With
${\nu}_{e}\;{\sim}\;10^{13}s^{-1}$, it requires
${\Gamma}$ as high as $10^{7}s^{-1}$. In principle, hopping of defects may
be present at zero temperature due to quantum tunneling, but the rate 
should be small. Thermal hopping of atomic defects in high-$T_{c}$ 
superconductors, at a sufficiently high temperature, can be much stronger. 
Assuming that  ${\Gamma}$
follows the Arrhenius law, ${\Gamma}={\nu}_{0}{\exp}(-U/T)$, with
${\nu}_{0}$ of order of the Debye frequency, $10^{13}s^{-1}$, we find that at
$77 K$ the barrier $U$ that would provide the critical hopping rate must be
about $1000 K$. Barriers as low as 870 K, presumably related to
hopping of isolated oxygen atoms in CuO planes, do exist in YBCO, see,
e.g., Ref. \cite{Gannelli} and references therein. In copper oxygen
superconductors without twin boundaries, BiSCCO in particular, oxygen 
vacancies can be 
the major source of pinning \cite{EMC90}. The significance of their 
hopping could be established, e.g., through measurements
of the magnetic relaxation induced by a sudden change in the pressure of
oxygen. If a significant relaxation is detected, it would be quite
conceivable that the phenomenon suggested in this paper could take 
place in the low-field high-temperature (large ${\Gamma}$) part of the 
phase diagram. The restoration of the long range order in the flux-line 
lattice due to the hopping of pinning centers would be effectively 
equivalent to the depinning of the lattice. It would result in the critical 
current, $j_{c}$, dropping to zero because $j_{c}$ goes down as a some power of 
$R_{t}$ \cite{Review}. Some additional experimental effort may be needed, 
however, to distinguish between the effect of hopping and the conventional 
thermal depinning or melting of the flux-line lattice. 

This work has been supported by the Department of Energy under 
Grant No. DE-FG02-93ER45487. 
\\


\begin{thebibliography}{99}
\bibitem{Remark} The generalization of all formulae to the case when 
this is not true is obvious.
\bibitem{damping} The limit of ${\eta}{\rightarrow}0$ is not 
described by our model because it is based upon a purely dissipative
dynamics.
\end{thebibliography}
\end{document}